\documentclass[amsfonts,showpacs,tightenlines,aps,12pt]{revtex4}
\usepackage{bm}

\newcommand{\bea}{\begin{eqnarray}}
\newcommand{\eea}{\end{eqnarray}}
 
\newcommand{\rhov}{\bm{\rho}}  
\begin{document}


\title{Microscopic substructure effects in potential-model descriptions 
of the $^7$Be$(p,\gamma)^8$B reaction} 
\author{Jutta Escher}\email{escher@triumf.ca}
\author{Byron K.~Jennings}\email{jennings@triumf.ca}
\affiliation{TRIUMF, 4004 Wesbrook Mall, Vancouver, BC, Canada V6T 2A3}
\date{\today}

\begin{abstract}
The spectroscopic factor arises from short-range effects in the nuclear wave
function.  On the other hand, cross sections for external capture reactions,
such as $^7$Be$(p,\gamma)^8$B at low energies, depend primarily on the
asymptotic normalization of the bound-state wave function -- a long-range
property.  We investigate the relationship between potential models and the
full many-body problem to illustrate how microscopic substructure effects
arise naturally in the relevant transition matrix element and can be (in part)
accounted for by a spectroscopic factor.
\end{abstract}


\maketitle

The $^7$Be(p,$\gamma$)$^8$B reaction at solar energies ($E_{cm}$ $\leq$ 20 keV)
plays an important role in the `solar neutrino puzzle' since the neutrino event
rate in the existing chlorine and water Cerenkov detectors is either dominated
by or almost entirely due to the high-energy neutrinos produced in the
subsequent $\beta$ decay of $^8$B.  Direct measurements of the
$^7$Be(p,$\gamma$)$^8$B rate, however, are extremely difficult, since the
absolute cross sections diminish exponentially at low energies.  Thus,
theoretical studies of this process become very valuable.  In addition to
microscopic theories, such as the nuclear shell model or cluster models,
potential models provide a popular framework for such investigations.  In the
latter approach, an average potential is used to generate single-particle wave
functions and to calculate various observables; microscopic substructure
effects are (partly) accounted for through the use of spectroscopic factors.
This strategy, however, has recently been called into question by
Cs\'{o}t\'{o}~\cite{Csoto00}, who argued that spectroscopic factors should not
be included in potential-model calculations of the $^7$Be$(p,\gamma)^8$B cross
section.

Cs\'{o}t\'{o}'s argument is based on the observation that the
$^7$Be$(p,\gamma)^8$B reaction at low energies depends only on the asymptotic
normalization of the $^8$B bound-state wave function~\cite{Muk99}, whereas the
spectroscopic factor arises from the short-ranged properties of the wave
function.  However, the unstated assumption underlying this line of reasoning,
namely that the short-range and long-range parts of the wave function are
independent, is incorrect.  For example, the bound state is normalized by $\int
d^{3N}\!  \bm{r} \, |\psi(\bm{r})|^2=1$, a global condition which illustrates
that a change in the wave function at any radius affects the normalization
integral and thus the value of the wave function everywhere, including in the
tail region.  Moreover, there exists a more subtle connection between the
asymptotic normalization and the short-range behavior of the wave function
based on the bound-state Lippmann-Schwinger equation~\cite{Jennings00}.  In
fact, the asymptotic normalization can be given in terms of an integral over
the interior of the nucleus~\cite{Jennings00}.

The arguments presented in ref.~\cite{Csoto00}, as well as the connection
between short-range and long-range parts of the wave function that was outlined
in ref.~\cite{Jennings00}, have motivated us to reconsider the proper treatment
of microscopic nuclear structure effects in potential-model descriptions of
external capture reactions.  In this paper we study how potential models are
related to the full nuclear many-body problem.  We set up a perturbative scheme
for the many-body system which, in the lowest-order approximation, gives the
potential model and discuss higher-order corrections coming from the
microscopic nuclear structure.  The spectroscopic factor will be shown to arise
naturally in this framework.

The eight-body wave functions for the $^8$B ground state and the proton-$^7$Be
scattering state are written as
$|\psi^B_8(\rhov_1,\rhov_2,\rhov_3,\rhov_4,\rhov_5,\rhov_6,\rhov_7)\rangle$ and
$|\psi^k_8(\rhov_1,\rhov_2,\rhov_3,\rhov_4,\rhov_5,\rhov_6,\rhov_7)\rangle$,
respectively.  Here $k$ is the asymptotic value of the momentum of the
scattering particle relative to the $^7$Be system, and we have introduced the
Jacobi coordinates
\begin{eqnarray}
\label{Eq_Jacobi}
\rhov_j \equiv \frac{\sum_{i=1}^j m_i \bm{r}_i}{\sum_{i=1}^j m_i} -
\bm{r}_{j+1} \;\; 
\mbox{for $j$=1,$\ldots$,8} \, ,
\end{eqnarray} 
with $\bm{r}_9 = 0$. The dependence of the wave functions on the center-of-mass
coordinate, $\bm{R}=\rhov_8$, can be factored out in the form of a plane wave
and will be suppressed throughout this paper.  

The eigenstates of the seven-body $^7$Be Hamiltonian,
$|\psi^n_7(\rhov_1,\rhov_2,\rhov_3,\rhov_4,\rhov_5,\rhov_6)\rangle$, 
form a complete set of wave functions when both bound and scattering states 
are included. 
Using this complete set, the eight-body bound-state wave function for $^8$B 
can be written as
\begin{eqnarray} 
|\psi^B_8(\rhov_1,\rhov_2,\rhov_3,\rhov_4,\rhov_5,\rhov_6,\rhov_7)
\rangle=\sum_{n=1}^\infty
|\psi^n_7(\rhov_1,\rhov_2,\rhov_3,\rhov_4,\rhov_5,\rhov_6)\rangle
\phi^n_B(\rhov_7) \, ,
\label{eq_expn} 
\end{eqnarray} 
where the summation symbol indicates a sum over the bound states and an
integral over the continuum states.  The expansion coefficients,
\begin{eqnarray}
\phi^n_B(\rhov_7)=\langle\psi^n_7(\rhov_1,\rhov_2,\rhov_3,\rhov_4,\rhov_5,\rhov_6)|
\psi^B_8(\rhov_1,\rhov_2,\rhov_3,\rhov_4,\rhov_5,\rhov_6,\rhov_7) \rangle \, ,
\end{eqnarray}
are obtained by inverting eq.~(\ref{eq_expn}).  Note that $\phi^1_B(\rhov_7)$
is the spectroscopic amplitude as defined in ref.~\cite{Csoto00}.  Its norm is
simply the spectroscopic factor, $S=\int d\rhov_7 |\phi^1_B(\rhov_7)|^2$ and we
will show below that $\phi^1_B(\rhov_7)$ can be identified as a one-body wave 
function.

The $^8$B Hamiltonian, $H_8$, can be written as
$H_8=H_7-\frac{\nabla^2_{\rhov7}}{2\mu_7}+\sum_{i=1}^7 V(|\bm{r}_i - \bm{r}_8|)$,
where $H_7$ is the Hamiltonian for the $^7$Be nucleus and $\mu_7 = m_8
\sum_{i=1}^7 m_i / \sum_{i=1}^8 m_i$ denotes the reduced mass of the eight-body
system.  Since $|\psi^n_7\rangle$ and $|\psi^B_8\rangle$ are eigenstates of
$H_7$ and $H_8$, respectively, we have:
\begin{eqnarray} 
\langle\psi^n_7|H_8| \psi^B_8 \rangle&=& E_8^B\langle\psi^n_7|\psi^B_8\rangle
          \nonumber \\ &=& E_7^n\langle\psi^n_7|\psi^B_8\rangle +
          \langle\psi^n_7|-\frac{\nabla^2_{\rhov7}}{2\mu_7}+\sum_{i=1}^7
          V(|\bm{r}_i - \bm{r}_8|)|\psi^B_8\rangle \; .
\end{eqnarray}
Inserting the expansion given in eq.~(\ref{eq_expn}) for $|\psi^B_8\rangle$, 
we obtain a set of exact, Schr\"odinger-like, coupled equations for the coefficients:
\begin{eqnarray}
(E_8^B-E_7^n)\phi^n_B(\rhov_7) = -\frac{\nabla^2_{\rhov7}}{2\mu_7}
\phi^n_B(\rhov_7) + \sum_{m=1}^\infty\langle\psi^n_7|\sum_{i=1}^7
V(|\bm{r}_i - \bm{r}_8|)|\psi^m_7\rangle \phi^m_B(\rhov_7)  .
\label{eq_schro}
\end{eqnarray}
The normalization condition for the $\phi^n_B(\rhov_7)$ follows from 
eq.~(\ref{eq_expn}): 
\begin{eqnarray}
\sum_{n=1}^\infty \int d^3\rhov_7|\phi^n_B(\rhov_7)|^2=1.
\label{eq_norm}
\end{eqnarray}
The spectroscopic factor is thus $S=1-\sum_{n=2}^\infty \int
d^3\rhov_7|\phi^n_B(\rhov_7)|^2$ and differs from one because of the $n\ne1$
terms in the sum.

From eq.~(\ref{eq_schro}) we can extract information on the long-range behavior
of the $^8$B ground state wave function.  The matrix element
$\langle\psi^n_7|\sum_{i=1}^7 V(|\bm{r}_i - \bm{r}_8|)|\psi^m_7\rangle$, for $n$ or
$m$ corresponding to a $^7$Be bound state, has a range on the order of the
potential convoluted with the bound-state wave function.  Thus it falls off
rapidly with increasing $\rhov_7$.  When both $\psi^n_7$ and $\psi^m_7$
describe continuum states, the matrix element has a long-ranged, but
infinitesimal, tail.  Solving eq.~(\ref{eq_schro}) outside the range of the
potential, we find that the $\phi^n_B(\rhov_7)$ decouple and fall off
exponentially, $\phi^n_B(\rhov_7) \propto \exp[-\sqrt{2 \mu_7 (E^n_7-E^B_8)} \,
\rho_7]$, where $\rho_7 = |\rhov_7|$.  Since $(E^n_7-E^B_8)$ is smallest for
$n=1$, the long-range behavior of the $^8$B wave function is dominated by
$\phi^1_B(\rhov_7)$ and the associated asymptotic normalization is given by
$A_n = \lim_{\rho_7 \rightarrow \infty} \phi^1_B(\rhov_7) \rho_7 \exp[+\sqrt{2
\mu_7 (E^n_7-E^B_8)} \, \rho_7 ]$.  Coulomb effects have been ignored in these
considerations.

In order to study the relationship between potential models and the full
many-body problem, we consider the above set of coupled differential equations
for arbitrary distances. However, we introduce a well-defined approximation
scheme. The off-diagonal matrix elements $V_{nm} \equiv
\langle\psi^n_7|\sum_{i=1}^7 V(|\bm{r}_i - \bm{r}_8|)|\psi^m_7\rangle$, $n \neq m$,
will be treated as a perturbation. In zeroth order we take $V_{nm} = 0$ for $n
\neq m$, that is, only the diagonal terms survive in eq.~(\ref{eq_schro}).  The
equations for the coefficients $\phi^n_B(\rhov_7)$ decouple and take the form
of Hartree equations for the eighth particle moving in the mean field of the
other nucleons.  The Hartree potential, $V_{nm}$, is different for each
$\phi^n_B(\rhov_7)$ since the two-body potential is convoluted over different
seven-body configurations.  Assuming that the ground state of $^8$B is built on
the ground state of $^7$Be, only $\phi^1_B(\rhov_7)$ can occur in the
zeroth-order expansion of $|\psi^B_8 \rangle$, that is, $|\psi^B_8 \rangle_0 =
|\psi^1_7 \rangle \phi^1_B(\rhov_7)$.  Coefficients $\phi^n_B(\rhov_7)$, with
$n>1$, are associated with excited states of the $^7$Be system and therefore
contribute (to lowest order) only to higher-energy states of $^8$B.  Since the
$^8$B states are orthonormal, it follows directly that the $\phi^n_B(\rhov_7)$
which correspond to bound states are normalized to one and those associated
with continuum states are normalized to a delta function.

From the considerations above, it becomes apparent that each $\phi^n_B(\rhov_7)$
is normalized and satisfies a single-particle, Schr\"odinger-like, wave
equation with a mean field of the Hartree type.  Therefore, the
$\phi^n_B(\rhov_7)$ can be identified with potential-model wave functions.  We
recall that $\phi^1_B(\rhov_7)$ is, by definition, the spectroscopic amplitude,
and thus find that the potential-model wave function associated with the ground
state of $^8$B yields the lowest-order approximation to the spectroscopic
amplitude.

Since, in lowest order, $\phi^n_B(\rhov_7) = 0$ for $n>1$, it follows from
eq.~(\ref{eq_schro}) that there is no first-order correction to the coefficient
$\phi^1_B(\rhov_7)$; the next correction to $\phi^1_B(\rhov_7)$ is of second
order in the off-diagonal coupling.  Nevertheless, coefficients
$\phi^n_B(\rhov_7)$ with $n>1$ contribute to the $^8$B ground state in first
order, thus the norm of $\phi^1_B(\rhov_7)$ changes from $S=1$ to $S<1$.  The
deviation of the spectroscopic factor from one is of second order in the
off-diagonal matrix elements $V_{nm}$.

The $^7$Be-p scattering state $|\psi^S_8\rangle$ can be expanded similarly to
the $^8$B bound state:
\begin{eqnarray}
|\psi^S_8(\rhov_1,\rhov_2,\rhov_3,\rhov_4,\rhov_5,\rhov_6,\rhov_7)
\rangle=\sum_{n=1}^\infty
|\psi^n_7(\rhov_1,\rhov_2,\rhov_3,\rhov_4,\rhov_5,\rhov_6)\rangle
\phi^n_S(\rhov_7) \, .
\label{eq_expn_scat}
\end{eqnarray}                    
Proceeding as before, we obtain a set of coupled equations for the expansion
coefficients $\phi^n_S(\rhov_7)$:
\begin{eqnarray}
 (E_8^S-E_7^n)\phi^n_S(\rhov_7) = -\frac{\nabla^2_{\rhov7}}{2\mu_7}
\phi^n_S(\rhov_7) + \sum_{m=1}^\infty\langle\psi^n_7|\sum_{i=1}^7
V(|\bm{r}_i - \bm{r}_8|)|\psi^m_7\rangle \phi^m_S(\rhov_7)
\label{eq_scatt}
\end{eqnarray}       
A perturbative treatment of the coupling matrix elements $V_{nm}$ is possible 
here as well: In lowest order the equations decouple and only
$\phi^1_S(\rhov_7)$ contributes to $|\psi^S_8\rangle$.  The $\phi^n_S(\rhov_7)$
with $n>1$ contribute to the scattering state in first order and the next
correction to $\phi^1_S(\rhov_7)$ is of second order in the $V_{nm}$. There is
no spectroscopic factor associated with the scattering wave function, since
$|\psi^S_8\rangle$ is normalized by its value at asymptotically large
distances, rather than by an integral.

Considering the long-range behavior of $|\psi^S_8\rangle$, we infer the
following from eq.~(\ref{eq_scatt}): If the incident proton has a kinetic
energy less than that of the first excited $^7$Be state ($E^2_7$=0.429
MeV~\cite{Tilley99}), $\phi^1_S(\rhov_7)$ becomes oscillatory for large
$\rhov_7$, while the $\phi^n_S(\rhov_7)$ with $n>1$ are exponentially damped.
Hence, for low reaction energies, the $n=1$ term dominates the long-range
behavior of the scattering wave function: $|\psi^S_8\rangle \rightarrow
|\psi^1_7\rangle \phi^1_S(\rhov_7)$.  At higher energies, $\phi^n_S(\rhov_7)$
with $n>1$ display oscillatory behavior and thus a coupled-channels approach to
the problem may become necessary.

The dominant contribution to the direct capture process $^7$Be(p,$\gamma$)$^8$B
arises from an electric dipole transition between the $^7$Be-p scattering state
and the $^8$B ground state~\cite{Kim87}.  The relevant matrix element is given
by:
\begin{eqnarray} 
\langle \psi^B_8|\sum_{i=1}^8 e_i \bm{r}_i | \psi^S_8 \rangle &=& \langle
\psi^B_8|\sum_{i=1}^7 \epsilon_i \rhov_i | \psi^S_8 \rangle\\ &=&
\epsilon_7 \sum_{n=1}^\infty \int d^3\rhov_7 \; \phi^{n*}_B(\rhov_7) \rhov_7
\phi^n_S(\rhov_7) \nonumber\\&&+ \sum_{{n=1}\atop{m=1}}^\infty \langle
\psi^n_7|\sum_{i=1}^6 \epsilon_i \rhov_i| \psi^m_7 \rangle \int d^3\rhov_7 \;
\phi^{n*}_B(\rhov_7) \phi^m_S(\rhov_7) \; ,
\label{eq_matel}
\end{eqnarray} 
where the ``Jacobi charges'', $\epsilon_i$ ($i=1,\ldots,8$), are defined such
that $\sum_{i=1}^8 e_i \bm{r}_i = \sum_{i=1}^8 \epsilon_i \rhov_i$ and
$\epsilon_8$ equals $Q$, the total charge of the system.  Since the final-state
wave function, $| \psi^B_8 \rangle$, is anti-symmetric and the dipole operator,
$\sum_{i=1}^8 e_i \bm{r}_i$, is symmetric, it is not necessary to
anti-symmetrize the initial-state wave function, $|\psi^S_8\rangle$.  Without
loss of generality, we can thus take the eighth nucleon to be the incident
particle.  Note also that the diagonal matrix elements in the last line,
$\langle \psi^n_7|\sum_{i=1}^6 \epsilon_i \rhov_i| \psi^n_7 \rangle$, vanish
since $\sum_{i=1}^6 \epsilon_i \rhov_i$ has odd parity and we can choose the
$^7$Be wave functions, $| \psi^n_7 \rangle$, to be states of good parity.

To lowest order in the perturbative expansion for the wave functions only terms
with $n=m=1$ survive. Thus we obtain:
\begin{eqnarray}
\langle \psi^B_8|\sum_{i=1}^8 e_i \bm{r_i} | \psi^S_8 \rangle \approx
\epsilon_7 \int d^3\rhov_7 \; \phi^{1*}_B(\rhov_7) \rhov_7 \phi^1_S(\rhov_7) \;
,
\label{eq_pert}  
\end{eqnarray}
where the Jacobi charge
\begin{eqnarray}
\epsilon_7=\mu_7\left(\frac{\sum_{i=1}^7 e_i}{\sum_{i=1}^7 m_i} -
\frac{e_8}{m_8}\right),
\label{mel}
\end{eqnarray}
accounts for the fact that the photon couples to both the initial nucleus and
to the incident proton~\cite{ChristyDuck61}.  The lowest-order term of the
perturbative expansion corresponds to the one-body approximation as given, for
example, in ref.~\cite{ChristyDuck61}.

The first-order corrections to the matrix element of the dipole operator
originate entirely from the $n=1$, $m\neq1$ and $n\neq1$, $m=1$ terms in the
second sum of eq.~(\ref{eq_matel}).  Note that $\langle \psi^1_7|\sum_{i=1}^6
\epsilon_i \rhov_i| \psi^m_7 \rangle$, for $m>1$, contains dipole excitations
of the $^7$Be core, that is, the transition operator acts on the internal
coordinates of the seven-body system.  These first-order effects are
short-ranged and will be suppressed at low energies, for which the reaction
takes place predominately in the tail of wave function.

While there are no first-order corrections to the first sum in
eq.~(\ref{eq_matel}), there are several second-order effects: First, there are
$n\neq1$ contributions, which will be suppressed at low energies since
$\phi^n_B(\rhov_7)$ and $\phi^n_S(\rhov_7)$ are exponentially damped for large
$\rhov_7$.  Second, there are corrections to both $\phi^1_B(\rhov_7)$ and
$\phi^1_S(\rhov_7)$ which are due to the off-diagonal coupling.  Since
$\phi^1_B(\rhov_7)$ and $\phi^1_S(\rhov_7)$ depend on only one coordinate,
these modifications can be accommodated by adjusting the one-body potential in
a manner similar to the construction of an optical-model potential.  Third,
$\phi^1_B(\rhov_7)$ is no longer normalized to one but to the spectroscopic
factor, as follows from eq.~(\ref{eq_norm}).  In addition, there are
second-order corrections to the second sum in eq.~(\ref{eq_matel}), such as
$n\neq1$, $m\neq1$ contributions.  These, again, correspond to dipole
excitations of the $^7$Be core and are thus short-range effects, which are
negligible at small energies.

From the above considerations it follows that the spectroscopic factor should
be explicitly included in potential-model descriptions of external capture
reactions.  When the off-diagonal matrix elements $V_{nm}$ are treated
perturbatively, the spectroscopic factor arises naturally in the normalization
of the bound-state wave function at second order and is present at all higher
orders.  For low-energy reactions first-order contributions and second-order
effects not contained in the spectroscopic factor are either small or may be
included by an appropriate modification of the one-body potential.  At higher
energies these first- and second-order corrections may become important and it
will no longer be sufficient to accommodate many-body correlations through the
use of spectroscopic factors alone.

The arguments for including the spectroscopic factor can be taken beyond
perturbation theory.  At low energies, eq.~(\ref{eq_pert}) will be valid more
generally since the neglected terms are associated with short-range effects
while the reaction occurs at large distances.  Thus we have to evaluate an
integral which contains two functions of one variable, a scattering wave
function, $\phi^1_S(\rhov_7)$, and a bound-state wave function,
$\phi^1_B(\rhov_7)$.  Formally, we can construct Schr\"odinger-like equations
for these coefficients using Green functions.  To illustrate this, we rewrite
the coupled differential equations for the expansion coefficients
$\phi^n_{\alpha}(\rhov_7)$ (eqs.~(\ref{eq_schro}) and~(\ref{eq_scatt})) as
follows:
\begin{eqnarray}
-\frac{\nabla^2_{\rhov7}}{2\mu_7} \phi^1_{\alpha}(\rhov_7) &+&
\langle\psi^1_7|\sum_{i=1}^7 V(|\bm{r}_i - \bm{r}_8|)|\psi^1_7\rangle
\phi^1_{\alpha}(\rhov_7)-(E_8^{\alpha}-E_7^1)\phi^1_{\alpha}(\rhov_7)
\nonumber \\
&=&-\sum_{m=2}\langle\psi^1_7|\sum_{i=1}^7
V(|\bm{r}_i - \bm{r}_8|)|\psi^m_7\rangle \phi^m_{\alpha}(\rhov_7)
\label{eq_alpha1} 
\end{eqnarray}
\begin{eqnarray}
-\frac{\nabla^2_{\rhov7}}{2\mu_7} \phi^n_{\alpha}(\rhov_7) &+&
\sum_{m=2}^\infty\langle\psi^n_7|\sum_{i=1}^7
V(|\bm{r}_i - \bm{r}_8|)|\psi^m_7\rangle
\phi^m_{\alpha}(\rhov_7)-(E_8^{\alpha}-E_7^n)\phi^n_{\alpha}(\rhov_7)
\nonumber \\
&=&-\langle\psi^n_7|\sum_{i=1}^7
V(|\bm{r}_i - \bm{r}_8|)|\psi^1_7\rangle \phi^1_{\alpha}(\rhov_7)
\label{eq_matrix}
\end{eqnarray}
Here $\alpha=B$ or $S$, for the bound and scattering states, respectively.
The latter equation is valid for $n>1$ and can be expressed as:
\begin{eqnarray}
\sum_{m=2}^{\infty} D_{nm}^{(\alpha)}(\rhov_7) \phi_{\alpha}^m(\rhov_7) 
&=& U_n^{(\alpha)}(\rhov_7) \; ,
\end{eqnarray}
where
\begin{eqnarray}
D_{nm}^{(\alpha)}(\rhov_7) &=& \left[ -\frac{\nabla^2_{\rhov7}}{2\mu_7} -
(E_8^{\alpha}-E_7^1) \right] \delta_{nm} + \langle\psi^n_7|\sum_{i=1}^7
V(|\bm{r}_i - \bm{r}_8|)|\psi^m_7\rangle \; , \nonumber \\ U_n^{(\alpha)}(\rhov_7)
&=& - \langle\psi^n_7|\sum_{i=1}^7 V(|\bm{r}_i - \bm{r}_8|)|\psi^1_7\rangle
\phi^1_{\alpha}(\rhov_7) \; .
\end{eqnarray} 
Let $G$ be the Green function matrix which solves $\sum_{k=2}^{\infty}
D_{nk}^{(\alpha)}(\rhov_7) G^{km}_{(\alpha)}(\rhov_7,\rhov_7')$
$=\delta(\rhov_7-\rhov_7')\delta_{nm}$.  It then follow that
\begin{eqnarray}
\phi^n_{\alpha}(\rhov_7) &=& -\int d^3\rhov'_7 \sum_{k=2}^\infty
G^{nk}_{(\alpha)}(\rhov_7,\rhov'_7) \langle \psi^k_7|\sum_{i=1}^6
V(|\bm{r}_i - \bm{r}_8|) | \psi^1_7 \rangle \phi^1_{\alpha}(\rhov'_7)
\end{eqnarray}           
is a solution of eq.~(\ref{eq_matrix}).  Inserting this result into the
right-hand side of eq.~(\ref{eq_alpha1}) yields an exact one-body,
Schr\"odinger-like, wave equation for $\phi^1_{\alpha}(\rhov_7)$.  For the
bound-state case, that is, for $\alpha=B$, this wave function is normalized to
the spectroscopic factor.

To summarize, we have studied the relationship between potential-model
descriptions of the capture reaction $^7$Be(p,$\gamma$)$^8$B and the full
many-body problem.  To this end, we set up a perturbative scheme for the
many-body system and identified the potential-model wave function as the
zeroth-order approximation to the spectroscopic amplitude.  We showed that the
lowest-order term in the perturbative expansion of the relevant transition
matrix element corresponds to the standard one-body approximation; the
spectroscopic factor was found to arise naturally in the normalization of the
bound-state wave function at second order.  At low energies, where first-order
effects and additional second-order corrections are small, the spectroscopic
factor allows one to incorporate the most important many-body correlations in
potential-model descriptions of external capture reactions and should thus be
included explicitly in the calculations.  At higher energies, where
contributions from the interior of the nucleus become important, additional
microscopic nuclear structure effects arise and including the spectroscopic
factor will no longer be sufficient.

\acknowledgements{This work was supported by funding from the Natural 
Sciences and Engineering Research Council of Canada.}


\begin{thebibliography}{2}

\bibitem{Csoto00}
A.~Csoto, Phys.~Rev.~C \textbf{61}, 37601 (2000).

\bibitem{Muk99}A.M.~Mukhamedzhanov and R.E.~Tribble, Phys. Rev. C
\textbf{59}, 3418 (1999); M.~Xu, C.A.~Gagliardi, R.E.~Tribble,
A.M.~Mukhamedzhanov, and N.K.~Timofeyuk, Phys. Rev. Lett. \textbf{73},
2027(1994).

\bibitem{Jennings00}
B.~K.~Jennings, nucl-th/9910005 (Phys.~Rev.~C, in press).

\bibitem{Tilley99}
D.~R.~Tilley, C.~M.~Cheves, J.~L.~Godwin, G.~M.~Hale, H.~M.~Hofmann, J.~H.~Kelley,
and H.~R.~Weller, ``Energy levels of light nuclei, A=7," preprint.  

   
\bibitem{Kim87}
K.~H.~Kim, M.~H.~Park, and B.~T.~Kim, Phys.~Rev.~C \textbf{35}, 363 (1987).

\bibitem{ChristyDuck61}
R.~F.~Christy and I.~Duck, Nucl.~Phys. \textbf{24}, 89 (1961).

\end{thebibliography}
\end{document}